\documentclass[epjST]{svjour}
\usepackage{graphics}
\usepackage{graphicx} 
\usepackage{hyperref}
\usepackage{url}
\usepackage{geometry}
\begin{document}
\title{Theoretical And Technological Building Blocks For An Innovation Accelerator}
\author{
	Frank van Harmelen\inst{1}\fnmsep\thanks{\email{frank.van.harmelen@gmail.com}} \and 
	George Kampis\inst{2,3} \and 
	Katy B\"orner\inst{4} \and
	Peter van den Besselaar\inst{5} \and
	Erik Schultes\inst{6,7} \and
	Carole Goble\inst{8} \and
	Paul Groth\inst{9} \and
	Barend Mons\inst{7,10,11,12} \and
	Stuart Anderson\inst{13} \and
	Stefan Decker\inst{14} \and
	Conor Hayes\inst{14} \and
	Thierry Buecheler\inst{15} \and
	Dirk Helbing\inst{16}
	}
\institute{
	AI Department, Division of Mathematics and Computer Science, Faculty of Sciences, VU University Amsterdam, de Boelelaan 1081a, 1081HV Amsterdam, The Netherlands \and 
	German Research Center for Artificial Intelligence, Trippstadter Strasse 122, D-67663 Kaiserslautern, Germany  \and
	E\"otv\"os Lorand University, Department of History and Philosophy of Science, Budapest, Hungary \and
	Cyberinfrastructure for Network Science Center, School of Library and Information Science, Indiana University 1320 E. 10th St, Wells Library, Bloomington, IN 47405, USA  \and
	Department of Organisation Sciences, Faculty of Social Sciences, VU University Amsterdam, The Netherlands  \and
	Department of Human Genetics, Leiden University Medical Center, Leiden, The Netherlands \and 
	Concept Web Alliance, Nijmegen, The Netherlands  \and
	Department of Computer Science, University of Manchester, Oxford Road, Manchester, M13 9PL, UK \and
	Knowledge Representation  \& Reasoning Group, Artificial Intelligence Section, Department of Computer Science, VU University of Amsterdam, De Boelelaan 1081a, 1081 HV  Amsterdam, The Netherlands \and
	Department of Human Genetics, Leiden University Medical Center, Leiden, The Netherlands \and
	Netherlands Bioinformatics Center, Nijmegen, The Netherlands \and
	Department of Medical Informatics, Erasmus Medical Centre, Rotterdam, The Netherlands \and
LFCS, School of Informatics, The University of Edinburgh \and
	Digital Enterprise Research Institute,  NUI Galway \and
	Artificial Intelligence Laboratory, Department of Informatics, University of Zurich, Andreasstrasse 15, 8050 Zurich, Switzerland \and
	Chair of Sociology, in particular of Modeling and Simulation, ETH ZŸrich, Clausiusstra§e 50, 8092 ZŸrich
	}
\abstract{
Modern science is a main driver of technological innovation. The efficiency of the scientific system is of key importance to ensure the competitiveness of a nation or region. However, the scientific system that we use today was devised centuries ago and is inadequate for our current ICT-based society: the peer review system encourages conservatism, journal publications are monolithic and slow, data is often not available to other scientists, and the independent validation of results is limited. The resulting scientific process is hence slow and sloppy. 
Building on the Innovation Accelerator paper by Helbing and Balietti \cite{helbing2011create}, this paper takes the initial global vision and reviews the theoretical and technological building blocks that can be used for implementing an innovation (in first place: science) accelerator platform driven by re-imagining the science system.  
The envisioned platform would rest on four pillars: (i) Redesign the incentive scheme to reduce behavior such as conservatism, herding and hyping; (ii) Advance scientific publications by breaking up the monolithic paper unit and introducing other building blocks such as data, tools, experiment workflows, resources; (iii) Use machine readable semantics for publications, debate structures, provenance etc. in order to include the computer as a partner in the scientific process, and (iv) Build an online platform for collaboration, including a network of trust and reputation among the different types of stakeholders in the scientific system: scientists, educators, funding agencies, policy makers, students and industrial innovators among others. 
Any such improvements to the scientific system must support the entire scientific process (unlike current tools that chop up the scientific process into disconnected pieces), must facilitate and encourage collaboration and interdisciplinarity (again unlike current tools), must facilitate the inclusion of intelligent computing in the scientific process, must facilitate not only the core scientific process, but also accommodate other stakeholders such science policy makers, industrial innovators, and the general public. 
We first describe the current state of the scientific system together with up to a dozen new key initiatives, including an analysis of the role of science as an innovation accelerator.  Our brief survey will show that there exist many separate ideas and concepts and diverse stand-alone demonstrator systems for different components of the ecosystem with many parts are still unexplored, and overall integration lacking. By analyzing a matrix of stakeholders vs. functionalities, we identify the required innovations. We (non-exhaustively) discuss a few of them: Publications that are meaningful to machines, innovative reviewing processes, data publication, workflow archiving and reuse, alternative impact metrics, tools for the detection of trends, community formation and emergence, as well as modular publications, citation objects and debate graphs.
To summarize, the core idea behind the Innovation Accelerator is to develop new incentive models, rules, and interaction mechanisms to stimulate true innovation, revolutionizing the way in which we create knowledge and disseminate information.
	} 
\maketitle
\restoregeometry
\pagebreak
\tableofcontents
\listoffigures
\pagebreak

\section{Introduction}
\label{intro}
Scientific research has been the driving force behind innovation, and it has been an enormously successful human endeavor. Virtually all major improvements in the quality of life of people around the world have been due to fundamental breakthroughs in scientific understanding. However, the system that we use today to organize collaboration, communication, competition and selection in modern science is essentially the same as it was devised in the 17th century, in the days of the first printed scientific journals delivered by horse or canal boats. There is a widespread feeling among scientists (and particularly, although not exclusively, among the younger generation), that the way we run science today is broken: the transition of scientific results into new social solutions (products, services) is measured in decades, the peer review system encourages conservatism, journal publications are large, monolithic and slow, data is often not available to other scientists, and the independent validation of results is limited. In short, the innovation process lacks coherence, agility and transparency. 

\subsection{Global Vision}
\label{sec:1}

The aim of the Innovation Accelerator \cite{helbing2011create} is to change this situation radically by creating new incentive models, rules, and interaction mechanisms to stimulate true innovation and new institutional designs to disseminate knowledge:

\begin{itemize}
\item Developing self-organizing reputation-based science platforms embedded in a self-balancing web of trust and ranking, that capture discipline specific features while supporting interdisciplinarity,
\item Creating new indices to discover high-quality work and new methods to analyze scientific productivity, facilitating early identification of innovations and trends
\item Inventing a corruption-proof reputation system that avoids tragedies of the commons in a globalizing world,
\item Designing new science forum, publication platform, and tools for large-scale cooperative projects based on insights from complex systems,
\item Devising customized automated recommendation platforms for stakeholders
\item Transforming publishers from gatekeepers of high quality information to innovation scouts and technology brokers,
\item From invention to innovation: Facilitating the creation of new business opportunities, markets and employment opportunities.
\end{itemize}

This paper discusses the theoretical foundation and technologies that can be applied to implement this grand vision by changing the science system. We start with a listing of news clippings on the current state of the scientific system, introduce the four main pillars of the proposed framework, and complete this section with desiderata.

\subsection{Current State of the Scientific System}
\label{sec:2}
The current state of affairs might be best described by facts and comments from major publications, publishing groups, Nobel laureates, and the press.

\vskip 10pt
\noindent
\textbf{Reliability of 'new drug target' claims called into question}
\par
\noindent
In a first-of-a-kind analysis of Bayer's internal efforts to validate 'new drug target' claims the company's in-house experimental data do not match literature claims in 65\% of target-validation projects, leading to project discontinuation.\par
\noindent \url{http://www.nature.com/nrd/journal/v10/n9/full/nrd3545.html}\par

\vskip 10pt
\noindent
\textbf{Spiraling cost of the peer review system}
\par
\noindent
"In 2008, a Research Information Network report estimated that the unpaid non-cash costs of peer review, undertaken in the main by academics, is £1.9 billion globally each year."\par
\noindent \url{http://www.publications.parliament.uk/pa/cm201012/cmselect/cmsctech/856/856.pdf}\par

\vskip 10pt
\noindent
\textbf{Unmanageable repositories of scientific publication}
\par
\noindent
Medline is now growing at the rate of one new research paper every minute.
\par
\noindent \url{http://www.nlm.nih.gov/bsd/medline_cit_counts_yr_pub.html}
\par

\vskip 10pt
\noindent
\textbf{Unintended consequences of the current peer review system}
\par
\noindent
"Either we all suck, or the system is broken" -- Jeffrey Naughton after mentioning that only one paper out of 350 submissions to SIGMOD 2010 received a unanimous ÒacceptÓ from its referees, and only four had an average accept recommendation.
\par
\noindent \url{http://lazowska.cs.washington.edu/naughtonicde.pdf}\par

\vskip 10pt
\noindent
\textbf{Protest against reviewing practice}
\par
\noindent
Journal editor of CCR produces empty issue to protest against the prevailing reviewing culture. \par
 \noindent\url{http://ccr.sigcomm.org/online/files/p3-v41n3ed-keshav-editorial.pdf}\par

\vskip 10pt
\noindent
\textbf{ÒOnly 47 papers (9\%) deposited full primary raw data online.Ó }
\par
\noindent
Public Availability of Published Research Data in High-Impact Journals.
Alsheikh-Ali AA, Qureshi W, Al-Mallah MH, Ioannidis JPA, 2011 \par
 \noindent PLoS ONE 6(9): e24357. \url{doi:10.1371/journal.pone.0024357}\par

\vskip 10pt
\noindent
\textbf{New Nature Group Editorial Policy}
\par 
\noindent
"The Nature Publishing Group will no longer accept submissions from humans (Homo sapiens)Ó because Òthe heuristics and biases inherent in human decision-making preclude them from conducting reliable science" \par
\noindent \url{http://www.nature.com/nature/journal/v477/n7363/full/477244a.html}\par

\vskip 10pt
\noindent
\textbf{Nobel Prize Winner coming out}
\par 
\noindent
The journals have long served as tombstones, certifications for tenure committees, rather than a forum in which ideas get argued. -- Nobel Prize Winner Paul Krugman\par
\noindent \url{http://krugman.blogs.nytimes.com/2011/10/18/our-blogs-ourselves/}\par

\vskip 10pt
These comments reflect the imperfect or broken socio-technical publication, incentive, and collaboration systems in existence in science today. Current practice does not serve the vivid exchange of innovative ideas; does not promote the sharing of data and methods; fails to motivate scientists and innovators by controversial peer reviewing practices; wastes money and energy on reviewing without producing reliable quality; puts the credibility of scientific results at risk for basing on wrong reputation mechanisms.
To further motivate these claims: a recent review \cite{alsheikh2011public} found that 44 of 50 leading scientific journals instructed their authors on sharing data, but fewer than 30 percent of the papers published followed the instructions. Opthof and Leydesdorff \cite{opthof2011comment} found evidence that peers find it extremely difficult (if not impossible) to distinguish between good and excellent research. The peer-review process becomes thus error-prone in such circumstances, and therefore it may be better to distribute funds (among the top group) than judging the value of research - a first selection between the tail and the top of the distribution may be less difficult than the actual calibration of value \cite{bornmann2010meta}.

Some aspects of the current practice are thus fundamentally flawed, and we suggest that a remedy should not address these aspects in isolation; rather, that the entire context of the process needs rethinking.

\subsection{Four Main Pillars}
\label{sec:3}
To address some of the above problems, the vision to improve and enhance the existing scientific system as outlined in this paper rests on the following pillars:

\vskip 10pt
\noindent
\textbf{(i ) Replace the PDF:} break up the current monolithic paper unit (which is, after all, only motivated by the limitations of the early printing press and distribution system), and introduce other research objects in the scientific exchange process, such as: 

\begin{itemize}
\item publishing data, 
\item publishing source code of programs used for simulations or data processing 
\item publishing experimental workflows, 
\item ÒpublishingÓ experimental resources to be used by other scientists, 
\item breaking up the current scientific paper into its separate constituents (motivation, background, hypothesis formulation, experimental design, data interpretation, conclusions, etc.) in support of algorithmic mining,
\item allowing all of these constituents to be referenced separately and interlinking them, thereby forming an entire new web of scientific debate.
\end{itemize}

\noindent
\textbf{(ii) Redesign the incentive scheme.} With these new Òresearch objectsÓ in the scientific discourse, new incentive schemes should be designed that exploit these new types of objects, and at the same time avoid current perverse incentives for conservatism, group herding, topic hyping, data hoarding, lack of independent verification, etc. 

\vskip 10pt
\noindent
\textbf{(iii) Use machine readable semantics.} Such new research objects should all be given machine readable formats: not only data, code and workflows, but also publications, debate structures, provenance information, trust and reputation metrics etc. should all be Òmachine interpretableÓ in support of large scale data mining and modeling. Similarly, computational research objects such as code and data should be Òmachine reproducibleÓ \cite{stodden2010scientific}. 

\vskip 10pt
\noindent
\textbf{(iv) Build online platforms for collaboration, including networks of trust and reputation} Such collaboration platforms and reputation networks will most likely be a combination of generic components and discipline specific elements. Together, the new objects for scientific exchange, governed by new incentive schemes and represented machine-interpretable formats should form the basis for online networks of trust and reputation to be used by all the different types of stakeholders in the scientific system: scientists, funding agencies, policy makers, students and industrial innovators, among others. 

\vskip 10pt
In this context the ÒqualityÓ of scientific artifacts is highly relevant (see also below, ÒDesiderataÓ). Quality, as trust and reputation, is subjective and built through interactions: research papers, even from the same author, have different impact on the scientific community, and they are assessed differently from diverse groups and stakeholders. For this reason, current research in the field of ranking on bipartite networks usually considers quality, trust and reputations together, and similar ideas should be implemented in new platforms. We emphasize that a reputation system on a network works by leveraging on limited and local human judgment power combined with collective networked filtering (cf. "Manifesto for the reputation society" \cite{masum2004manifesto}). Such a platform, modeled by a bipartite user-object network, may represent a candidate working ground for personalized recommender systems which are extremely helpful in delivering the right content/expertise to the right stakeholder.

\subsection {Implementation}
The implementation of the four pillars will require the combination of IT technology (to technically facilitate the new and user-friendly ways of science), complex systems mathematics (to model and analyze the current system in order to understand why it is failing/succeeding, to study and design optimal incentives and to optimize innovation), and social science (because after all science is a social system in the first place). It requires a deep understanding of the social and socio-technical mechanisms in which science and innovation are embedded, see Section~\ref{sec:In} on Models of Innovation. Existing technological building blocks have to be identified, advanced, and combined to arrive at a system architecture and functionality that is greater than the sum of its parts (see Section~\ref{sec:tech}). Successful design and implementation as well as wide-spread adoption of the Innovation Accelerator will depend on a detailed understanding of stakeholder needs and an implementation of functionality that truly addresses these needs, see Section~\ref{sec:stake}. Ultimately, the envisioned Innovation Accelerator will impact science, competitiveness, and society at large, see Section~\ref{sec:imp}.

\section{Desiderata}
\label{sec:4}
In order to succeed and experience wide-spread adoption, the envisioned Innovation Accelerator has to satisfy a number of desired properties, some of which we briefly outline here: 

\vskip 10pt
\noindent
\textbf{Support for the entire scientific process.} Much of the technology currently in daily use in scientific laboratories and on the desktops of scientists chop up the scientific workflow in disjoint parts, with little or no support for smoothly switching and migrating from the on-line network of scientific publications to hypothesis formation, or for moving data between experimental design and the laboratory workbench, or from the laboratory workbench to the actual publication. All these gaps in the scientific process must be bridged by Òmental copy-pasteÓ of teams of scientists, and much knowledge and information (both quality and provenance) gets lost in the process. 

\vskip 10pt
Actual scientific creation is characterized by an extended lifecycle that begins with the conception and documentation of an idea, proceeds via the identification of relevant partners and publications, the carrying out of various procedures often resulting in models and data, and leads to the finished end product. Conceived in this complex way, the latter does not need to be static but can be involved in various feedback loops leading to model and post-publication amendments.

Many innovative systems that can individually support elements of the scientific process from the original idea conception to collaborative skywriting to innovative forms of publishing have been proposed and realized. These need to grow into a comprehensive system with a coherent vision to automate and integrate all aspects of the scientific production system, i.e., the scientific process itself. 

\vskip 10pt
\noindent
\textbf{Support for collaborative science and flexible interdisciplinary publishing.} The scientific social process of hypothesis generation, experimental validation, communication and dissemination follows old patterns, dictated by technology of the printing press and communication at the speed of the postal system. These are characterized by: individualism instead of collective collaboration, linear chains instead of networks, static publishing instead of dynamic, and the appearance of herding effects and the lack of rich context.
What we need, by contrast, is true interdisciplinary collaboration, communication based on adaptive and dynamic filters, and a layered, agile, distributed and context-rich publication model. The involvement of community tools in the collaboration process would much benefit from social media tools (such as social tagging, trust networks) in the scientific workflow.

\vskip 10pt
\noindent
\textbf{An online network of dynamics, trust and reputation.} Current ranking systems that drive careers and reputations in science (and are therefore important elements of the current incentive system for scientists) are one-dimensional, and only count simple metrics like impact factors, H-index citation-statistics, or accumulated grant-acquisition. Instead, new indicators should be developed \cite{helbing2011create} that rank both people, projects and publications on a multi-dimensional scale, which facilitate the identification of quality, trust \cite{cattuto2009collective} and the detection of new trends and dynamics. 

\vskip 10pt
\noindent
\textbf{Augmenting the scientific process using computers.} Semantic annotation of the text of scientific papers; recording the provenance of datasets; formally representing of the central claims of a scientific paper; making explicit the structure of the debate in a scientific community; algorithms for calculating reputation: many of the technologies and approaches discussed in this paper make it possible to include computers as substantial ÒpartnersÓ in the scientific production process. Automated reasoning is no longer a luxury, but is absolutely necessary to the future of the scientific enterprise as the rate of data generation already vastly exceeds human reasoning capability. It can be argued that without machine assisted reasoning and inference, continued data generation becomes a catastrophic public and private waste. 

For example, with increased degrees of formalization of the scientific communication it is possible to do things like automatic inconsistency detection (two different models deriving different conclusions from the same data).

\vskip 10pt
\noindent
\textbf{Support for all stakeholders in the scientific process.} Any improved scientific system should facilitate not only the core scientific process, but also accommodate other stakeholders. 
Science is currently elitist and closed, whereas the advent of the Internet opens the context to the public including various stakeholders. Using new tools, they could get directly involved.

We envision impact on a broad range of stakeholders: of course individual scientists and their teams, but also publishers (journals, editors), scientific communities (represented, e.g., by learned societies), institutional bodies (universities, government agencies, grant-giving organizations), but equally students, policy makers and the general public.  For example, real-time observability of a trusted network of the dynamics in science should help policy makers to decide where to strengthen funding, and should help industrial innovators deciding which communities to turn to, and where to place investments in technological development based on new scientific findings. This support plays a key role in fulfilling the role of accelerating innovation. 

\section{Models of Innovation}
\label{sec:In}
\begin{figure}
	\centering
	 \includegraphics[width=0.99\linewidth]{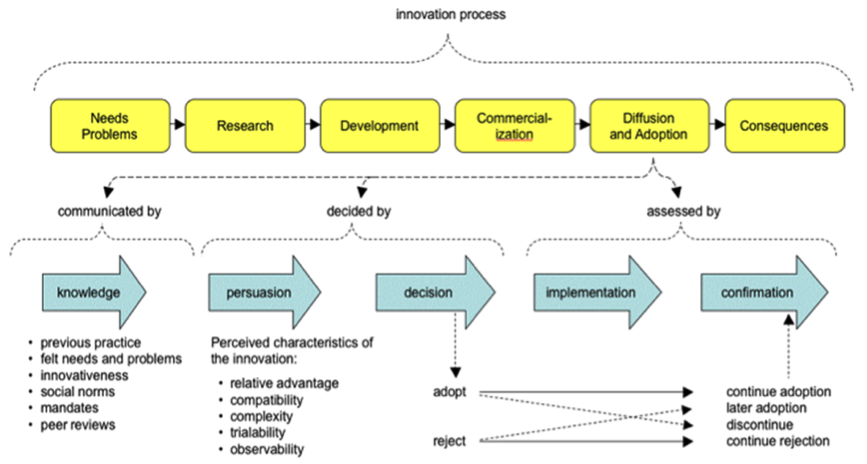}
	 \caption{Innovation as a linear system}
	\label{fig:lin_innov}
\end{figure}

Prior work \cite{kline1986overview,nelson1977search} aims to identify the distinct actors and processes that lead from research to technological innovation. Early models emphasized the linear succession of different steps, see Figure~\ref{fig:lin_innov}.

Recent work tries to capture the innovation process as feedback system of flows that support innovation and cooperation. The model by Cara\c{c}a \emph{et al.} \cite{caracca2009changing} is shown in Figure 2. It was developed for product and service innovation in established enterprises but can be reasonably easily adapted to take account of spin-outs or start-ups.\footnote{The differences in the models for start-ups and spinouts would be in terms of proximity and origin of the product in the development and refinement model.  For start-ups one might believe the company would have a close connection with the market, have little in the way of organisational knowledge and have a loos connection with the science knowledge stream.  For spinouts, one might believe that there would be a close connection with the science knowledge with less information on market and business.  These considerations would condition the starting point in the development cycle.}  

The model has at its core a 5-stage product or service development process.  This process is connected to three key knowledge channels via interfaces. Each knowledge channel is represented in the same way -- as an established body of knowledge that is being added to by a continuing research process. Here we argue that the innovation accelerator is concerned with improving the quality of access to the identified knowledge streams and evolving the product development cycle to take account of development in the way the knowledge streams deliver.  

In their conclusion \cite{caracca2009changing} emphasize the Òmulti-channelÓ nature of their model, arguing that the Òscience channelÓ is but one amongst others:

\begin{quote}
\emph{ÒScience remains a fundamental source of innovation, but in a plural knowledge context characterized by a multiplicity of intertwining channels where cooperation and technical information flows abound and take the form of learning processes. In order to transform the knowledge produced by R\&D into commercial results, firms need to engage in interactive learning externally with customers and markets, and manage the feedbacks from the broader social and institutional environment. Today in the networked globalizing learning economy the internal interaction across specialized functions continues to be equally important. Therefore, we have introduced a multi-channel learning model where research aiming at understanding markets and organizations appears on an equal footing with scientific research aiming at developing new technology and where experience-based learning is recognized as a prerequisite for transforming scientific knowledge into economic performance.Ó}
\end{quote}

\noindent
We may note, using information from other sections of this paper that the above Òscience channelÓ is actually not a single channel but is rather a complex Òmulti-channelÓ entity and the interface between channels and to the innovation processes for products and services is not a simple query interface but rather involves complex social and technological interactions. In addition, the science channel is undergoing very rapid revision and reworking, and could be the source of innovation for many other ÒknowledgeÓ channels.

\begin{figure}
	\centering
	 \includegraphics[width=0.99\linewidth]{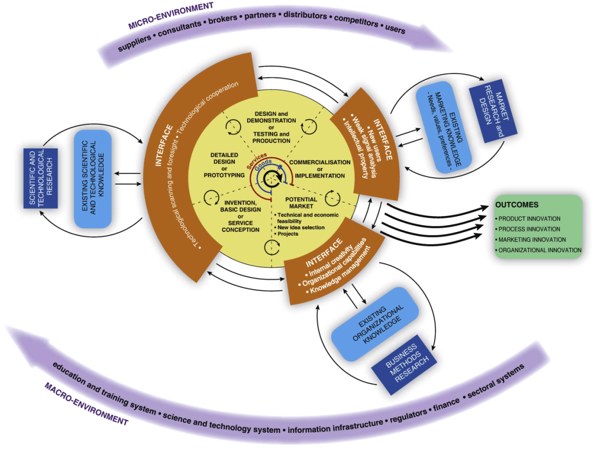}
	 \caption{Innovation as a cyclic system}
	\label{fig:cyc_innov}
\end{figure}

In characterizing the interface between the science channel and the product \& service development processes we can detect three main elements that interact to provide a range of possible modes of engagement between the science knowledge channel and the product \& service development process:

\vskip 10pt
\begin{enumerate}
\item \textbf{Discovery:} this involves finding the data, publications, individuals, communities, and live data streams that could provide the knowledge platform for a particular group of innovations.  This will be difficult because it almost certainly will involve working across disciplines, identifying possibly highly specialized niches within each discipline and translating work across disciplines to build a potentially useful configuration of contacts, results, data and processes that provide the starting point for innovation.
\item  \textbf{Fusion:} this is the process of investigating if the configuration identified in discovery could provide the answers to the sorts of questions the innovators are interested in.  This will involve some research inside the enterprise and could result in significant interactions with the researchers to understand the significance of their work and possibly to sponsor additional applied research.
\item  \textbf{Recontextualising:} the work of transferring knowledge from the academic context into that of the enterprise can be extremely difficult  since it is difficult to determine in advance what elements in the academic context might be essential to the effectiveness of a process.
\end{enumerate}

\vskip 10pt
Each of these elements plays an important role in each of the other main knowledge channels identified in the innovation model and it is clear that many of the approaches being adopted in science have the potential to benefit other channels and other channels have practices that could be beneficial in the scientific setting:

\vskip 10pt
\noindent \textbf{Market knowledge:}

\begin{itemize}
\item Discovery involves identifying the key communities that both comprise and have an important contribution to make to understanding the market. This is combined with relevant live data streams and data collection. This discovery process might involve the combination of evolving data sets that indicate candidate individual market participants combined with tools to confirm their potential membership of the group. The capacity to detect emerging scientific communities is similar to the capacity to identify a globally distributed niche market and its associated commentators, reviewers and opinion formers.
\item Fusion involves bringing together crowdsourced evidence together with other metrics to help in the formation of a new market and help establish the community as a responsive resource to help identify requirements and evolve the product in response to changing circumstances.
\item Recontextualisation involves understanding how discovered features of the populations across many different cultures and circumstances translate into marketing messages that are either universal across the population or can be translated back into messages that achieve their intention in localized contexts.
\end{itemize}

\noindent \textbf{Business knowledge:}
\begin{itemize}
\item Discovery: increasingly, large organizations are using open innovation techniques to identify people in the organisation as potential problem solvers and as holders of valuable knowledge about the processes of the enterprise. Conscious building of communities that orient to particular problem domains and the development of reputation and trust systems within these networks play an increasing role in solving complex production or logistics issues.
\item Fusion: for some time companies have been interested in structuring and accessing their organizational memory, see \cite{d2004inside} for a range of examples of how companies shape and integrate process, production and market knowledge to adapt and respond the changing demand for their products.
\item Recontextualisation: plays a particularly critical role in business environments.  It is reasonably well known the cultural factors can often undermine so called Òjust in timeÓ production processes and accommodating the move from a US setting to a European setting where the ISO 9000 series of standards require different instrumentation and reporting g on processes point to the specificity of business knowledge and the need to identify critical elements in the success of an approach before attempting to transfer the process. Much of this is very similar to considering how to move an experimental process from one lab to another.
\end{itemize}

It is clear that there are many appealing analogies between scientific knowledge channels and other knowledge channels involved in innovation processes. It is also clear that much work is required to put an appealing analogy on a firmer footing. The goal of this section is just to argue for a view of innovation systems as developing particular forms of knowledge and so supply a particularly valuable test bed for many of the ideas and components that could be developed within the proposed FuturICT programme. In particular, the use of the Planetary Nervous System as a means to enable global reach in identifying individuals and communities that constitute markets; the global participatory platform as a means of integrating data and tools to discover and fuse diverse knowledge channels and the living earth simulator as a source of models with which to interpret knowledge and to simulate the development of markets. We envisage investigating:

\vskip 10pt
\begin{enumerate}
\item The evolving structure of the knowledge streams identified in the innovation system diagram.  In the discussion on the innovations in the science knowledge channel detailed elsewhere in this paper it is clear that the fine structure both of the established scientific knowledge base and the process of creating new science is evolving rapidly.  The diagram depicted in Figure~\ref{fig:innov_sys}
is a first attempt to draw out the fine structure of that channel in order better to see where innovations in scientific knowledge have impact and how this might impact the ÒexternalÓ interface to companies and other ÒconsumersÓ of scientific work.  The diagram indicates that scientific research is highly differentiated (and different cultures are dominant in different disciplines and sub-disciplines, see Knorr-Cetina \cite{knorr1999epistemic}) and these cultures rest on an existing knowledge base that is increasingly actively curated using tools and techniques to decompose and repurpose scientific work with varying degrees of success.  This curation work involves complex configurations of tools, people and data and just how to support this activity in a global, open platform is a very significant challenge. One small example is a project carried out in Edinburgh using named entity recognition techniques drawn from the Natural Language processing community to connect gene sequencing data to the literature on protein-protein interaction \cite{alex2008automating}. This is one relatively small, commercially-driven, example -- we can expect to see much larger, more complex, attempts to do this kind of repurposing as European science become more focused on the Societal Challenges incorporated in the Horizon 2020 vision of the European Commission.

\begin{figure}
	\centering
	 \includegraphics[width=0.7\linewidth]{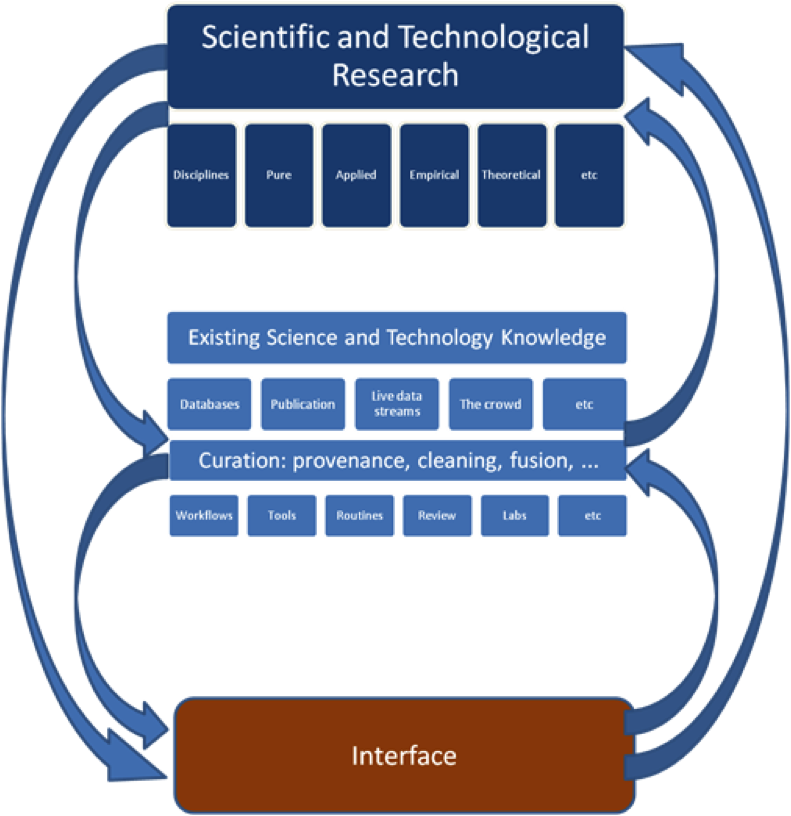}
	 \caption{Innovation as a cyclic system}
	\label{fig:innov_sys}
\end{figure}

\item The extent to which there are similarities between the three knowledge streams and to what extent there might be synergy in considering the transfer of methods used in one knowledge stream to another.  Here there are clear starting points in the areas of using crowdsourcing (possibly with filtering for relevant expertise), reputation systems, trend detection and the capacity to find and develop emerging communities. There are probably many other areas of common interest. Each area needs detailed work to see how the specific circumstances of a channel conditions the way tools and data are used.
\item The definition of the interfaces is a key feature of the model. As the knowledge production process evolves and is consciously re-formed by ICT we need to consider what kinds of interaction the interfaces should support. In the innovation system diagram, the identification of interfaces is helpful but there is limited exploration of the nature of the interface. Since the interfaces should support rich types of human interaction it will need to support features like reputation and trust while supporting very diverse types of access to the knowledge channel. Empirical study of the sorts of demands such interfaces are expected to deal with and whether there is a useful single interface or a family of Òdomain specificÓ interfaces is a better approach will also require the study of current innovation systems and the likely evolution of such systems.
\item The extent to which the change in the nature of modern products and services changes the relationship between the knowledge streams and the product. Many modern products and services are reflexive, in the sense that they carry a model of their environment and their impact on the environment. This provides the basis for the use of products as ÒprobesÓ into markets and into how they are developed and deployed. This offers new approaches to acquiring both market and production knowledge. More work will be required to investigate the impact of these types of products and services and whether their arrival needs further revision of innovation system models.
\item Whether there are any key resources that have been omitted or marginalized in the description of the innovation process. One clear omission is the inclusion of a distinct ÒdesignÓ channel. The natural approach is to include this in the business knowledge stream but conceivable this limits the role of design to the production process when it is clear that it can be critical to the creation of ÒpremiumÓ products (Apple being the leading example of design-led innovation).  
\end{enumerate}

The goal of this section was to sketch the potential lines of investigation that might be followed in the construction of an Innovation Accelerator that takes the innovations in approaches to the creation of Scientific Knowledge as a starting point. We do not explicitly consider the impact of success in this enterprise. The potential impact is to empower the whole of industry to become more innovative from a single-handed entrepreneur with massively enhanced access to knowledge channels to global corporations who already possess more knowledge than they can manage using current techniques resulting in many exploitable ideas languishing in research labs.

\section{Technological Building Blocks}
\label{sec:tech}
This section surveys existing databases, tools and services, and collaboration platforms that can be readily adopted and extended for use in the Innovation Accelerator. 

\vskip 10pt
\subsection{Scholarly Datasets}
There are a number of efforts that aim to make science and technology data available in easy to use digital formats. Among them are commercial data providers such as Thomson Reuters or Elsevier, government data providers, and academic efforts. Google Scholar is a notable example since currently it is the largest paper database in the world, however its data is presently not available for a direct database access (only interactively via Google's interface), thus it is not ãopen dataÓ. Here we exemplarily review open data efforts that aim to make massive amounts of paper, patent, funding, and government data available for general and direct usage.

\vskip 10pt
\noindent \textbf{Scholarly Database}
The Scholarly Database (SDB) at Indiana University \cite{la2007scholarly} aims to serve researchers and practitioners interested in the analysis, modeling, and visualization of large-scale scholarly datasets. The online interface at \url{http://sdb.cns.iu.edu} provides access to four datasets: MEDLINE papers, U.S. Patent and Trademark Office patents (USPTO), National Science Foundation (NSF) funding, and National Institutes of Health (NIH) funding -- over 25 million records in total. Users can register for free to cross-search these databases and to download result sets as dumps, in ready to use formats such as tables, co-author networks, or patent citation networks,  for scientometrics research and science policy practice. The SDB wiki (\url{http://sdb.wiki.cns.iu.edu}) has detailed information.

\begin{figure}
	\centering
	 \includegraphics[width=0.99\linewidth]{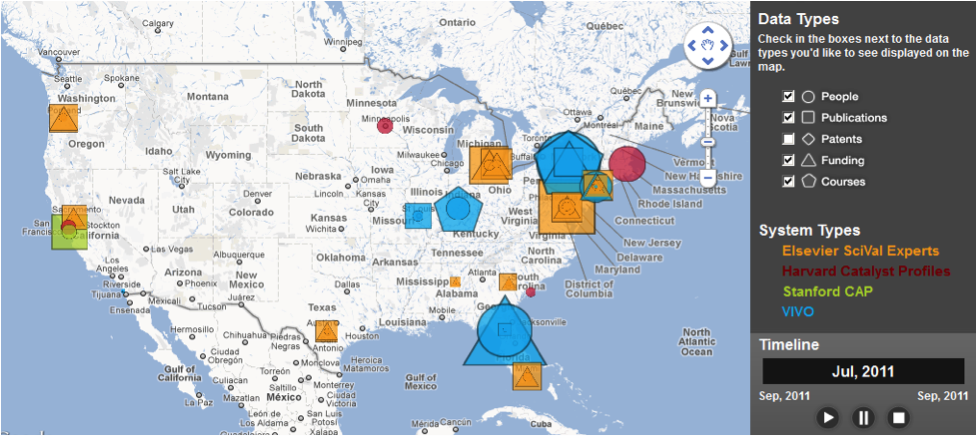}
	 \caption{Geolocations of institutions that implemented different (color-coded) researcher networking systems. Data holdings of different record types (denoted marker shapes) and number of records (area size) are indicated and updated monthly}
	\label{fig:resnet}
\end{figure}

\vskip 10pt
\noindent  \textbf{VIVO Researcher Network}
The VIVO International Researcher Network effort (\url{http://vivoweb.org}) aims to facilitate scholarly collaboration, research discovery, and better research evaluation as highly demanded by individual researchers and administrators of research institutions \cite{borner2012}. VIVO is an open source Semantic Web application that using data about teaching, research, and service activities from institutional repositories and publication data from commercial providers to support formerly time-consuming and mission-impossible searches, analyses, and insights. Originally developed and implemented at Cornell University, recent NIH funding in the amount of \$12 million converted VIVO into an enterprise application in support of research and scholarship that is being adopted by many schools in the USA, and organizations in Australia and China, as well as (more recently) in the Netherlands and Hungary. In interactive interface of different researcher networking systems and their data holdings can be found at \url{http://nrn.cns.iu.edu}, see Figure~\ref{fig:resnet}.

\begin{figure}
	\centering
	 \includegraphics[width=0.8\linewidth]{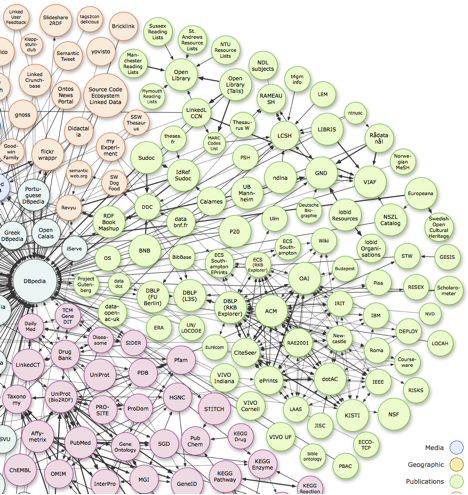}
	 \caption{Section of the Linked Open Data cloud devoted to scientific publications}
	\label{fig:linked}
\end{figure}

\vskip 10pt
\noindent \textbf{Linked Open Data}
The Linked Open Data cloud is a collection of data sources that have been published and interlinked using Semantic Web principles \cite{bechhofer2008semantic}: designating entities by URIÕs, declaring the types of these entities in RDF Schema or OWL, and expressing relations between these entities in RDF. At the time of writing, the Linked Data Cloud (LOD) contains some 25 billion relations between billions of objects. A significant portion of that distributed web-based datastructure is devoted to scientific bibliographic information (Figure~\ref{fig:linked}), 
ranging from the institutional VIVO endpoints described above to entire bibliographic databases for a single field (e.g., DBLP, covering much of Computer Science, listing over 600.000 authors), commercial digital libraries (such as the ACM), to collections of more informal scientific information such as SlideShare.net. Interestingly, the LOD collection directly addresses a number of the requirements we listed above: itÕs not tied to a single discipline, it can form the basis for an online reputation network, and by virtue of RDF, RDF Schema and OWL, it provides machine accessible semantics.

Linked Data is already providing a platform for an observatory of scientific practice and discourse by linking various sources of information: scientific publication indexes, blogs, news, open-source software repositories, social network information about scientists and their communities. This provides the ability to integrate and analyze data at large scale from a variety of sources in order to support hypotheses about scientific community building and impact, the roles of individuals, and also emergent topics. Linked Data and accompanying techniques support the analysis of relations between different data sets and domains, an automated identification of correlations and semi-automatic linkage and disambiguation. Together with state-of-the-art crawling technologies, this will enable the building of datasets that capture scientific practice and discourse represented on (micro-)blogs, news posts, Websites, but also scientific databases and software repositories. Furthermore, use of semantic technology will enable citation networks based on and ÒtypedÓ citations and claims, such as support, contradiction, etc. Complex system analysis and social network analysis on this data will provide the analytical background for understanding lifecycles and dynamics of scientific communities and discourse - the discourse leaders, the sub-community splits and merges, the strengthening of ties of between communities and even the emergence of Kuhnian paradigm shifts \cite{belak2011life}.

Several other open data initiatives exist and cannot be listed here -- just to name a few, the DataVerse (\url{thedata.org}), \url{DataCatalogs.org}, The Open Knowledge Foundation (\url{okfn.org}) and of course several open scientific repositories like \url{ArXiv.org} or \url{DataDryad.org}. Another example of interest is the Econophysics Forum (\url{http://unifr.ch/econophysics}) as an example of an online scientific community, sharing different kinds of content (papers, reviews, blogs, events, and so on).

Other platforms that facilitate the exchange of scientific (bibliometric) data include Mendeley, iamScientist, MS Academic, and again Google Scholar, with advanced derivatives such as Scholarometer (\url{http://scholarometer.indiana.edu}) or Harzing's ÒPublish or PerishÓ (\url{http://www.harzing.com/pop.htm}).

\subsection{Platforms for Code-Reuse, Data Mining, and Visualization}
As an example of a generic data and code-reuse and analysis platform, we discuss CIShell. 

The Cyberinfrastructure Shell (CIShell) (\url{http://cishell.org}) (Herr II et al., 2007) provides an easy to use, modular, scalable means to integrate and use datasets, algorithms, tools, and computing resources. It builds upon the Open Services Gateway Initiative (OSGi) (\url{http://www.osgi.org}) specification, thus leveraging a large amount of industry standard code and know-how. Today, CIShell is at the core of the major cyberinfrastructures and tools that serve the Information Visualization (\url{http://iv.cns.iu.edu}), Network Science (\url{http://nwb. cns.iu.edu}), Science Policy (\url{http://sci2.cns.iu.edu}), and Epidemics research communities. 

\begin{figure}
	\centering
	 \includegraphics[width=0.9\linewidth]{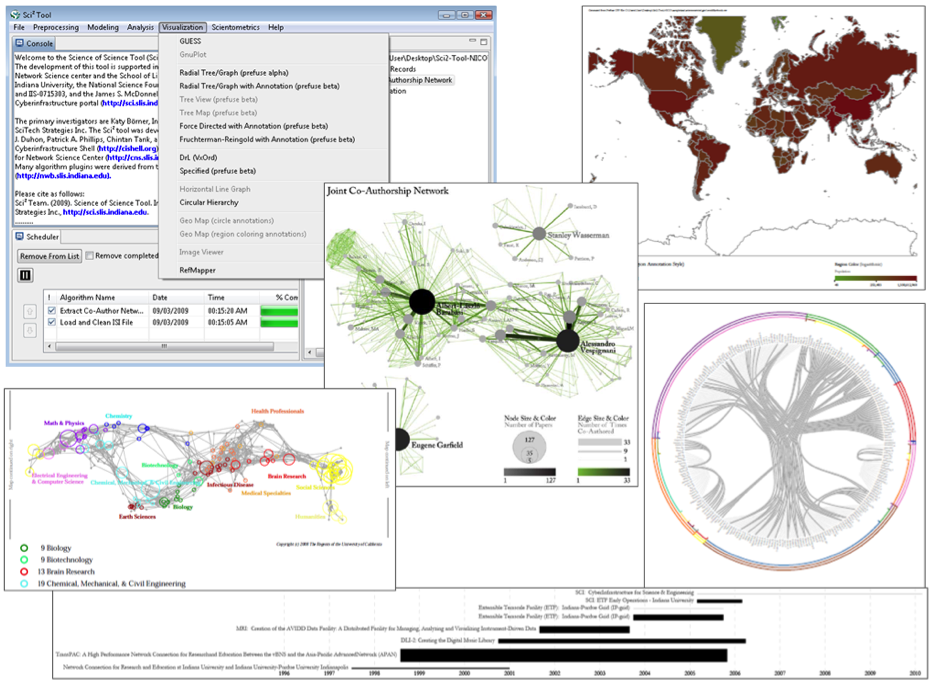}
	 \caption{Plug-and-play macroscope interface}
	\label{fig:cishell}
\end{figure}

Since 2008, a number of other teams have adopted OSGi and two EU framework projects (\url{www.textrend.org} and \url{www.dynanets.org}) have adopted CIShell. Among the OSGi adopters are Cytoscape, Taverna, MAEviz, and TEXTrend. Cytoscape (\url{http://www.cytoscape.org}) led by Trey Ideker, UCSD, is an open source bioinformatics software platform for visualizing molecular interaction networks and integrating these interactions with gene expression profiles and other state data \cite{shannon2003cytoscape}. Taverna Workbench (\url{http://taverna.sourceforge.net}) led by Carol Goble, University of Manchester, UK is a free software tool for designing and executing workflows (Hull et al., 2006). Taverna allows users to integrate many different software tools, including over several thousand web services now published in the sciences. MAEviz managed by Shawn Hampton, NCSA (\url{https://wiki.ncsa.uiuc.edu/display/MAE/Home}), is an open-source, extensible software platform which supports seismic risk assessment based on the Mid-America Earthquake Center research \cite{elnashai2008overview}. 

OSGi/CIShell has been adopted by the EU projects TEXTrend and Dynanets. TEXTrend (\url{http://www.textrend.org}) led by George Kampis, E\"otv\"os University, Hungary develops a framework for the easy and flexible integration, configuration, and extension of plugin-based components (see Figure~\ref{fig:cishell}).
in support of natural language processing (NLP), classification/ mining, and graph algorithms for the analysis of business and governmental text corpuses with an inherently temporal component \cite{kampis2009dynamic}; DynaNets (\url{http://www.dynanets.org}) coordinated by Peter M.A. Sloot at the University of Amsterdam, The Netherlands develops algorithms to study evolving networks; SISOB (\url{http://sisob.lcc.uma.es}) an Observatory for Science in Society Based in Social Models. 

As the functionality of OSGi-based software frameworks improves and the number and diversity of dataset and algorithm plug-ins increases, the capabilities of custom tools will expand.

Note that data-driven science is not only happening in the natural sciences. Science as a whole is affected. Increasingly, the humanities are also being influenced by the availability of large volumes of research data. Large amounts of data from the domain of qualitative humanities and social science research have become available for quantitative analysis. Private enterprises (Google Books and Earth, YouTube, Flickr, Twitter, Freebase , IMDB, among others) as well as public and non-profit institutions (Europeana, Wikipedia, DBPedia, Project Gutenberg, WordNet, Perseus, etc.) are in the process of collecting, digitizing, and structuring vast amounts of humanities-specific information (among other things), and creating technologies, applications, and services (Open Calais, Amazon's Mechanical Turk, ReCaptcha, ManyEyes, etc.), which are transforming the way research is done in the humanities, as much as it is transforming the natural sciences. The deployment of (often deep) computer analytics on these corpora extends humanities research to questions that are well beyond what can currently be accomplished by human beings. 

\subsection{Scientific Workflow as Part of the Publications}
A scientific workflow in scientific problem-solving is the process of combining data and processes into a configurable, structured and repeatable sequence of steps. Scientific workflow systems are finally becoming an adopted mechanism for the (reusable) encoding and (repeated) execution of computational scientific methods \cite{goble2009impact,ludascher2009scientific}. As illustration, in the IEEE e-Science Conference in 2011, two thirds of the published papers referred to workflows. 

For example, the Taverna scientific workflow management system enables: accessing to data, computational and analytic services; the assembling data processing and analysis pipelines; and recording the provenance of the yielded computed \emph{in silico} experimental results. 

From a scientific publication point of view, workflows and other computational methods highlight two major innovations:

\vskip 10pt
\noindent \textbf{Publishing method:} Rather than just publishing the results of method, novel platforms such as QScience (\url{http://www.qlectives.eu/qscience}), HubZero (\url{http://hubzero.org}) and myExperiment (\url{http://www.myexperiment.org}) enable the social networking and sharing of scientific methods, allow scientists to share and execute scientific workflows, or to find and reuse publicly available workflows, in the case of myExperiment, or simulation scripts in the case of HubZero. These support scientific communication and collaboration at a distance, and provide a new forum for computational method publication.

\vskip 10pt
\noindent \textbf{Executable publications:} Increasingly, publishers are open to experimentation with digital methods that go beyond the traditional publishing of a static report, either on paper or as PDF. Ideally, a scientific publication could be conceived as a complete record of the scientific production process: documentation of method in a reproducible way (software or experiment specification), of data archiving and processing, up to complex machine-readable documents. We quote from the website of ElsevierÕs 2011 ÒExecutable paperÓ challenge (\url{http://www.executablepapers.com}): 

\begin{quote}
\emph{"Data sets, code, and software are but some of the crucial elements in data intensive research; yet, these elements are noticeably absent when the research is recorded and preserved in perpetuity by way of a scholarly journal article. Further, most researchers do not deposit data related to their research article; and if they do so, it is often deposited on their personal or institutional websites, lacking consistency, reliable dissemination, discoverability, proper association (to the research article), documentation, validation, and preservation. To address all these concerns and to accommodate the ever increasing body of data intensive science, considerable adaptations to the existing journal article are fundamental to accommodating the need to disseminate, validate, and archive research data, as well as a method to allow this data, in some way or form, to be validated, citable, tractable, and executable. To achieve this adaptation to scholarly publication, several issues must be addressed; the most vital being: executability, long term preservation, validation of data and code, intellectual property rights and provenance."}
\end{quote}

A proposed example solution is the Collage Authoring Environment \cite{nowakowski2011collage} which amplifies and extends several already existing high-end computational expert tools (such as the Sweave tool \cite{leisch2002sweave} that puts R statistical computing scripts into LateX documents), by offering an intuitive user interface than can be operated by the average scientist and supports methods and data publishing in mainstream scientific papers. Tools such as UTOPIA Documents (\url{www.getutopia.com}) provide a stepping stone between traditional PDF and executable PDFs through mechanisms that make a PDF ÒactiveÓ and its contents executable or linked to underlying datasets.

Workflow publishing tools support scientific collaboration, knowledge acceleration and method transparency through the harnessing of human computation, and workflow systems support collaboration, acceleration and transparency through automated computation.

\section{User Needs and Proposed Functionality}
\label {sec:stake}
In this section we look beyond the demonstrator systems that we discussed above, and look at the technological innovations and indeed paradigmatic shifts that are needed to realize the vision outlined in the earlier sections of the paper. We start with an analysis of the different stakeholders in the science and technology system, the different tasks that these stakeholders perform, and the different information needs that come with these different tasks. 

\subsection{Stakeholders, Their Tasks, and Information Needs}
Different stakeholders in the S\&T system have vastly different needs and perform a large number of different tasks. Here we exemplarily review the needs of major stakeholders and detail the diverse roles scholars have.  

\vskip 10 pt
\begin{itemize}
\item Funding Agencies -- Need to monitor (long-term) money flow and research developments, identify areas for future development, stimulate new research areas, evaluate funding strategies for different programs, decide on project durations, funding patterns.
\item Scholars- - Want easy access to research results, datasets, relevant funding programs and their success rates, potential collaborators, competitors, related projects as well as publications (research push).
\item Industry -- Is interested in fast and easy access to major results, experts, etc. It influences the direction of research by entering information or running competitions/challenges on needed technologies (industry-pull).
\item Publishers -- Need easy to use interfaces to massive amounts of interlinked data. Need to communicate data provenance, quality, and context.
\item Society -- Needs easy access to scientific knowledge and expertise.
\end{itemize}

A closer look at scholars quickly reveals that most of them perform diverse roles with distinct information requirements such as:

\vskip 10 pt
\begin{itemize}
\item Researchers and Authors -- need to select promising research topics, students, collaborators, and publication venues to increase their reputation. They benefit from a global view of competencies, reputation and connectivity of scholars; hot and cold research topics and bursts of activity, and funding available per research area.  
\item Editors -- have to determine editorial board members, assign papers to reviewers, and ultimately accept or reject papers. Editors need to know the position of their journals in the evolving world of science. They need to advertise their journals appropriately and attract high-quality submissions, which will in turn increase the journalÕs reputation and lead to higher quality submissions. 
\item Reviewers -- read, critique, and suggest changes to help improve the quality of papers and funding proposals. They need to identify related works that should be cited or complementary skills that authors might consider when selecting project collaborators.
\item Teachers -- teach classes, train doctoral students, and supervise postdoctoral researchers. They need to identify key works, experts, and examples relevant to a topic area and teach them in the context of global science. 
\item Inventors -- create intellectual property and obtain patents, thus needing to navigate and make sense of research spaces as well as intellectual property spaces.  
\item Investigators -- scholars acquire funding to support students, hire staff, purchase equipment, or attend conferences.  Here, research interests and proposals have to be matched with existing federal and commercial funding opportunities, possible industry collaborators and sponsors. 
\item Team Leads and Science Administrators -- many scholars direct multiple research projects simultaneously. Some have full-time staff, research scientists, and technicians in their laboratories and centers. Leaders need to evaluate performance and provide references for current or previous members; report the progress of different projects to funding agencies. 
\end{itemize}

A detailed user and task analysis also reveals number of tasks that are common to different stakeholders and tasks. Among them are: 

\vskip 10 pt
\begin{itemize}
\item Finding experts: as (paper/grant) reviewers, panelists or speakers, or to hire them.
\item Evaluating scholars: individual scholars, teams, departments, or entire institutions in terms of number of papers, citations, grant funding, etc.
\item Communicating impact: in terms of number of papers, citations, grant funding, etc.
\item Strategic resource allocation: What (time/funding) investment will have the most impact (higher citation counts, external funding intake, and profit).
\end{itemize}

\subsection{Functionality Specification}
Given the above user needs, a number of system requirements and functionalities can be identified. Among them are: 

\subsubsection{Publications, that are meaningful to machines}
As stated above, automated reasoning becomes necessary as the rate of data generation exceeds human reasoning capability. As Fox and Hendler \cite{fox2009semantic} write: \emph{"[science] will generate petabytes of data that must be analyzed by hundreds of scientists working in multiple countries and speaking many different languages. The digital or electronic facilitation of science, or eScience, is now essential and becoming widespread."}

The Nature Publishing Group has brought this point to the public attention in a humorous editorial stating that \emph{"Nature will no longer accept submissions from humans (Homo sapiens)Ó because Òthe heuristics and biases inherent in human decision-making preclude them from conducting reliable science"}: 
\url{http://www.nature.com/nature/journal/v477/n7363/full/477244a.html}. 

Nanopublications (\url{http://www.nanopub.org}) have been proposed as a single format for research objects \cite{groth2010studying,mons2008calling}. A nanopublication is the smallest unit of publishable information, and has the form of a semantic assertion (minimally Subject-Predicate-Object) plus provenance metadata (minimally author, time-stamp). When appropriately serialized (using the RDF mark-up language), nanopublications can be machine readable, interoperable, and retrieved using advanced semantic web query methods. 

Nanopublications are designed to expose heterogeneous Big Data ensuring rapid dissemination, interoperability, and long-term persistence. Because nanopublications permit any data to be attributed to its authors, to institutions and to specific projects (like government funded grants) their scientific impact can be tracked, creating incentives for the exposure of both legacy and forthcoming data in nanopublication format. Tools for authoring, sharing, interpreting, citing, tracking and prioritizing nanopublications are currently under development as part of the Open PHACTS consortium (\url{http://openphacts.org}), which also provides the first sandbox environment for scalability and stress tests. Implicit in the design of a nanopublication infrastructure is a range of semantic technologies including ontologies and curated vocabularies, services resolving ambiguities and locating appropriate URIs, and tools for locating, creating and mapping relations between terms and concepts.

\subsubsection{Alternative Reviewing Process}
The lead quote cited from Jeffrey Naughton \emph{"Either we all suck or something is broken"} sets the context for this section. It was widely believed by scientists, policy makers and the general public earlier that the peer review system is the gold standard quality assurance method in science. It is becoming increasingly clear that this is not the case. 

Anecdotal historical evidence, as well a number of systematic social experiments (e.g. \cite{baxt1998who})  show the fundamental incapability of the current review system to uncover errors in submitted manuscripts. Furthermore, the (mostly hidden) costs of this system are significant: a 2008 Research Information Network report (UK House of Commons, Science and Technology Committee 2011 \cite{house2011}) estimated that the global costs of (mostly unpaid) peer review undertaken by academics, is 1.9 billion pounds annually. Some are now advocating that the current academic review practice (reviews by certain deadline, written by a small number of anonymous reviewers, who receive no particular award) be replaced by mechanisms more akin to Òscientific crowd sourcingÓ, or voting systems, or more longitudinal emergent processes. The authors of Altmetrics.org are claiming ÒInstead of waiting months for two opinions, an articleÕs impact might be assessed by thousands of conversations and bookmarks in a weekÓ. To some extent the physics community has already been moving in this direction with the widespread adoption of the \url{arXiv.org} system. 

What communication systems, publication platforms, award schemes, and social structures would be needed to sustain such an alternative reviewing process? Can we predict in advance (and consequently guard against) some of the disadvantages that would certainly come with na•ve versions of such new schemes? These questions definitely need further study. But in general there can be little doubt that relying on more information should be better than relying on less. This is exactly what the alternative reviewing procedures envisage.
For instance, as we speak, nanopublications have become foundational to driving real-time stakeholder observations of the scientific process. Along with Nature Publishing Group, LUMC (the Leiden University Medical Center) is developing a Òfutures marketÓ, including a Nanopublication Futures Bet Book where experts can record their estimated odds on a nanopublication being true. This crowd-sourced annotation will identify individual nanopublications from the sea of knowledge as high priority candidates needing further verification (e.g., experimental confirmation in the case of hypothetical nanopublications). Rather than a relatively small number of policy makers deciding on national funding priorities (using arduous and often closet methods), a nanopublication futures market would allow the community of experts to vote on and observe real-time ranking of scientific assertions. Nanopublications that surface as high priority will motivate capable labs to test assertions likely to be of interest to the community.

\subsubsection{Data Publication} 
Besides the peer review system, the maxim of independent reproducibility of results is a major cornerstone that is assumed to be ensuring the stability, quality and strength of the scientific edifice. But as with the peer review system, this mechanism is increasingly under pressure in the practice of every science. 

For a start, researchers are not rewarded for reproducing previous experiments, as reproducing experiments are not as popular with editorial boards as ground breaking new results. However, the importance of reproducibility cannot be questioned: in experimental particle physics, it is customary to have (at least) two different collaborations working on the same experiment on the same accelerator. Eventually only one of them will achieve fame for a discovery, but the work of the other collaboration(s) provides validation to the results or might stir debates on the procedures adopted and/or possible misconduct.  Furthermore, even when a scientist is inclined to try and reproduce results, it is often impossible to actually do so. Publications rarely contain all the information an independent researcher would need in order to reproduce the experiment (e.g. \cite{prinz2011believe}). Neither the data nor the experimental process (ÒworkflowÓ) is typically reported in a sufficiently detailed way to allow reproducing the experiment. It is not even in the individual scientistÕs interest to facilitate replication of their results: publishing the data is often seen as Ògiving away the competitive advantageÓ. There is currently a hot debate on data accessibility. Proprietary limit the opportunities to most scholars, not only towards the validation of existing results, but also towards the formulation of alternative approaches which could open new research avenues and opportunities.  Quoting Bernardo Huberman: \emph{"If another set of data does not validate results obtained with private data, how do we know if it is because they are not universal or the authors made a mistake?"} \cite{huberman2012sociology}.

What community practices, social norms, publishing platforms, data-formats, workflow description languages and execution engines would we need in order to substantially increase the reproducibility and actual reproduction (and hence increased validation) of scientific claims?

\subsubsection{Workflow Publication and Reuse} 
The preparation, integration and analysis of data at the scales of modern science requires scalable processing methods.  A scientific workflow in scientific problem-solving is the process of combining data and processes into a configurable, structured sequence of steps. Computational scientific workflows implement semi-automated solutions that are: 

\begin {itemize}
\item a systematic and automated way of processing data pipelines across incompatible data sets and integrating analytical applications; 
\item a way of capturing that process so that results can be reproduced, the method accurately reviewed and validated and know-how shared, reused and adapted; 
\item a visual scripting interface so that computational scientists can assemble pipelines and access services shielded from low-level programming concerns; and 
\item an integration and access platform for the growing pool of independent resource providers that avoids the scientist having to download and learn their codes. 
\end {itemize}

Workflows have the potential to liberate scientists from the drudgery of routine data processing so they can concentrate on scientific discovery. They shoulder the burden of routine tasks; they represent the computational protocols needed to undertake data-centric science; and they open up the use of processes and data resources to a much wider group of scientists and scientific application developers.  The importance of maintaining an accurate record of workflows grows. First, creating workflows requires expertise that is hard won and often outside the skill set of the researcher. Hence there is significant benefit in establishing shared collections of workflows that contain standard processing pipelines for immediate reuse or for repurposing in whole or in part. These aggregations of expertise and resources can help propagate techniques and best practices. Second, workflows are methods, which, alongside materials are a crux of the scientific approach. Without method, results are not reproducible or open to validation. Thus workflows are first class citizens, to be exchanged and published as data and articles are. 

Naturally, such computational workflows need machines readable formats in order to be machine executed. The workflow formats choreograph the control and data flow between the steps and also produce a log of the execution so that results might be audited. Currently no universal model exists for workflows: the many in-the-field scientific workflow management systems vary in their computational models, the kinds of components they execute and the data types they support. One size does not fit all. However, a shared provenance model and a common service-based approach would go some way to workflow interoperability. 

The myExperiment (\url{http://www.myexperiment.org}) social web site has demonstrated that by adopting content-sharing tools for repositories of workflows, we can enable social networking around workflows and provide community support for social tagging, comments, ratings and recommendations, social network analysis and reuse mining (what is used with what, for what, and by whom), and mixing of new workflows with those previously deposited. Currently, myExperiment supports the workflows of 21 different workflow systems.

\subsubsection{Alternative Metrics} 
Citation scores such as the H-index have become the almost universal tool by which to measure the quality of work of a scientist. It would be a remarkable feat of dimensionality reduction if such a complex agent (a scientist), performing such a complex task (scientific research) could be reliably measured on just a 1-dimensional scale. Nevertheless, tenure committees and funding agencies alike seem to be optimistically relying on such one-dimensional metrics, despite its well-known limits (it depends on the age and the discipline of the scholars, so it is not adequate to compare scientists and/or departments). The popularity of the H-index has led to a massive adaptation process by scholars, whose publication and citation habits aim at maximizing this quantity, to the detriment of creative scientific work. The H-index is also based on a rather narrow view of the scientific process and its outcome: scientific publications are the only currency of interest, and citation count is the only metric for this currency. What about other scientific outputs such as experimental data (and how much the data is being re-used), reviewing the work of others, maintaining scientific infrastructure (a non-trivial amount of scientific software development is done by Ph.D. students and postdocs), organizing scientific events. We are in need of other, richer metrics that do more justice to the contributions that a scientist can make to the advancement of science beyond measuring citation scores on journal papers. 

The Altmetrics manifesto addresses these concerns by calling for (and right away suggesting) a set of powerful alternative tools for scientific performance evaluation. Early speculation regarding altmetrics by Tarabonelli, Neylon, Priem and others \cite{taraborelli2008soft,priem2010altmetrics,neylon2009article,priem2010scientometrics} is beginning to yield to empirical investigation and working tools: for instance, Priem and Costello \cite{priem2010and} and Groth and Gurney \cite{groth2010studying} find citation on Twitter and blogs respectively. ReaderMeter (\url{http://readermeter.org/}) computes impact indicators from readership in reference management systems. Total Impact (\url{http://total-impact.org}) collates social media metrics from 15 sources. Datacite (\url{http://datacite.org}) promotes metrics for datasets. Future work as envisaged in this paper must continue along these lines.

\subsubsection{Scientific Discourse and Social Media} 
Scientific discourse is traditionally being played out by setting out and describing goals in conference and project grant announcements, reporting on performed experiments and obtained experimental results, reviewing of publications, summarizing the state of the art in text books and the popular press, etc. Most of this information is communicated by means of natural language, in addition to formal descriptions of methods, data sets and experimental settings. Novel forms of scientific communication that emerge and must be integrated in an Innovation Accelerator include tweets and blogs, that are particular forms of nano-publications or micro-contributions (\url{http://www.communitywiki.org/en/MicroContribution}).

It is a fact of reality that scientific discourse is increasingly taking place in online social platforms, both before ÒtraditionalÓ publications appear and after. For example, in mathematics, the PolyMath project (\url{http://polymathprojects.org/}) gathers hundreds of mathematicians throughout the world to work on advanced mathematical proofs on a wiki. In computer science, many community events such as the TREC information retrieval challenge (\url{http://en.wikipedia.org/wiki/Text_Retrieval_Conference}), the trading agent competition (\url{www.sics.se/tac/}), and the ontology alignment competition are organized completely online (\url{http://oaei.ontologymatching.org/}). In the Life Sciences, on-line data repositories like UniProt ((\url{www.uniprot.org/}) are central to the field, but additionally sites like BioStar (\url{http://biostar.stackexchange.com/}) act as a clearinghouse for sharing expertise. While the Web thus provides us with new traces of emergent science practice and communities of practice that are up-to-date and deep, the information is distributed, diffuse and messy. In order to be useful for tracking and predicting interesting emergent phenomena (such as new theories and new trans-disciplinary communities), this data must be integrated, refined, condensed, and cleaned. This not only requires the application of information integration techniques, but also the understanding of the semantics of the data such that quality analysis and inference can be carried out.

Along similar lines, we further mention the European project Liquid Publications (\url{ http://liquidpub.org/}). The project aimed at changing the way scientific knowledge is produced, disseminated, evaluated, and consumed. A practical results which can be useful to the IA is Liquid Books, a novel evolutionary model for writing books that mix the benefits of multi-author collaborations with the agility, freedom and simplicity of personal editions.
Social media sites like Twitter are already extensively used to spread scientific information. Recent studies 
\cite{letierce2010using,letierce2010understanding} studying twitter streams of scientific conferences show that twitter streams provide means to detect trend topics of the event, by (1) combining the amount of tweets posted with the conference hashtags and (2) studying URLs, other hashtags and retweets. In addition, it is possible to detect hubs and authorities. However, analyzing current tagging habits of scientists on Twitter revealed that the way users tag content leads mainly to messages targeted to peer researchers, while other communities could be interested in what they are talking about. The Social Semantic Web -- the enrichment of the social Web with Semantic Web information \cite{breslin2009social} enables a much finer grained support for scientific discourse. 

Semantic microblogging platforms \cite{passant2010rethinking} enable users to provide more fine granular information by making more information from microblogs machine-readable as part of the Linked Open Data Cloud (using specialized vocabularies such as SIOC and FOAF). By integrating and representing a formal representation of the scientific discourse \cite{passant2009swan}, this discourse becomes available for study as Linked Data. SWAN/SIOC provides an argumentative ontology \cite{passant2009swan,groza2009abstract}, currently in use for representing research discussions about AlzheimerÕs and Parkinson's diseases. Groza \emph{et al.} \cite{groza2007salt1,groza2007salt2,groza2008adding,groza2009abstract,groza2011capturing} developed a semantic-technology based claim federation architecture for externalizing the discourse structure of scientific papers, using Linked Open Data. Similarly Groth \emph{et al.} \cite{groth2010anatomy} have proposed reinventing scientific communication as a web of interconnecting nanopublication -- i.e. essentially claims with evidence, which also use Semantic Web technology.

\subsubsection{Crowdfunding for Science} 
ÒCrowdfundingÓ is a funding scheme to complement traditional funding for research at different maturity stages \cite{agrawal2011geography,brown2011}. It addresses several issues found in traditional funding (see, e.g., \cite{wu2010tackling,putnam2001bowling}) from ÒinventionsÓ all the way to ÒinnovationsÓ \cite{nelson1977search,nelson1982evolutionary,schumpeter1934theory} and leverages all sources from knowledge supply chains while following demand and needs from different communities (society, investors, markets). In this sense, crowdfunding may importantly contribute to the Innovation Accelerator. In particular, the approach enables a move towards breadth (supporting research that is not following current research trends) and involves more stakeholders in the process of asking and investigating upon research questions. We expect new herd effects to enable research Òoutside the coreÓ and point to issues that are not in the focus of large funding institutions yet relevant for a globalizing society \cite{buecheler2011}. 

In the case of crowdfunding, the objective is to collect money for investment; this is generally done by using social networks, in particular through the Internet (Twitter, Facebook,  LinkedIn and different other specialized blogs). In other words, instead of raising the money from a very small group of sophisticated investors, the idea of crowdfunding is to obtain  it from a large audience (the "crowd"), where each individual will provide a very small amount. This can take the form of equity purchase, loan, donation or pre-ordering of the product to be produced \cite{belleflamme2010crowdfunding}. An empirical investigation of important properties of crowdfunding has been given in \cite{belleflamme2010crowdfunding}. Hence, in the spirit of "Research 2.0" \cite{shneiderman2008copernican}, research projects that do not fit the current research trends or other patterns may benefit from the collective wisdom (or needs) of the masses by receiving funds that large institutions would not provide. This system builds on a true meritocracy, judged by the value to a large group of individuals breaking through barriers of political (or other) considerations of the usual decision makers. This, in turn, leads to an automatic prioritization of projects (committing funds acts as a weighted vote). For projects approaching market maturity, the usefulness is even more evident: receiving a monetary support from a community likely evidences the existence of potential customers of the innovation.

The practical implementation of such a crowdfunding approach is straightforward: Independent of the advancement of an inquiry (early idea stage or approaching market maturity), research projects can present themselves on a platform (stating the total amount of money that is needed to reach the desired state: the Òtarget amountÓ), competing for attention and micro-donations. Interested individuals can browse this platform to find projects they would like to support and commit a small (or large) amount to one or several projects, based on needs or even simpler motives like pure interest. As soon as the target amount is committed by the crowd, the project is launched and the committed money is collected through a micropayment platform (like PayPal or a credit card processing system). The money is only collected once the target amount has been reached by a common commitment of crowd members (an alternative approach is to put the committed funds on a blocked account until the target amount is reached). 

In order to initiate such a platform and reach a critical mass, a set of research questions that have not been in the focus of the funding community and/or are of general interest needs to be accumulated. In addition, an open board or interface should be provided where researchers can post and present their planned projects and court for micro-funds. Lastly, a suitable form of communication between members of this community must be secured to enable clarifying (and potentially enriching) discussion and a form of Òcollaborative filteringÓ \cite{su2009survey,shneiderman2008copernican} of ideas. 

In a second phase, tools for co-creation or crowdsourcing in general should be provided such that community members can not only contribute money, but also collective intelligence and manpower (similar to AmazonÕs ÒMechanical TurkÓ \cite{kittur2008crowdsourcing}, thus fully enabling  ÒCitizen ScienceÓ in the sense of Irwin \cite{irwin1995citizen}).

\subsubsection{Trend Detection, Community Formation, Emergent Science Fields} 
Besides in career decisions, citation scores also play a crucial role in how to filter and prioritize what to read and not to read in the growing mountain of scientific publications \cite{hull2008defrosting,renear2009strategic}.

For illustration: ArXiv is growing at the rate of 5000 papers a month, Medline at 10x that speed, at 50.000+ papers a month, amounting to a new paper every minute, PubMed now contains around 18 million papers. Citation scores are a current tool of the trade to do such filtering. Initiatives like AltMetrics mentioned above \cite{priem2010altmetrics} propose alternative mechanisms for filtering the literature, based on mechanisms much more akin to social networks (e.g., Mendeley) and social media (scientific blogging and microblogging). 

\begin{figure}
	\centering
	 \includegraphics[width=0.7\linewidth]{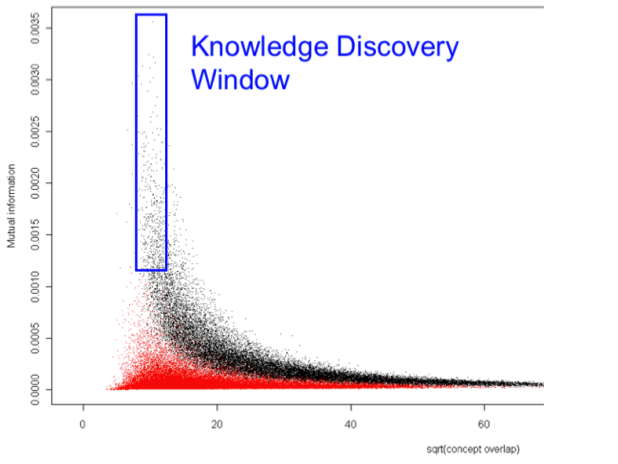}
	 \caption{Phase transition in mutual information of implied semantic links}
	\label{fig:kndwindow}
\end{figure}

As another example, at LUMC (the Leiden University Medical Center mentioned earlier), a technology called Òconcept profilesÓ have been applied to large textual and other data sets, allowing explicit and implicit semantic links to be identified. By filtering networks of implicit information to a critical point (essentially a percolation phase transition on the semantic network), it is possible to identify a deeply buried Ôknowledge discovery windowÕ that is invisible to ordinary statistical analysis. Concept profile data mining can be ongoing and run in real-time, allowing the limits of knowledge discovery to be mapped and the forecasting of new discoveries, see Figure~\ref{fig:kndwindow}.

Topical communities could also be identified by means of graph clustering techniques applied to the semantic networks above as well as to citation and collaboration networks of papers and their authors. In particular, it is crucial to monitor the evolution in time of the scientific landscape. The identification of dynamic clusters will enable us to keep track of the birth, evolution and death of topics. It will permit the early identification of pioneers of new fields, and the determination of when a community is formed, which could help to timely set up dedicated venues and funding schemes, along with career incentives for promising newcomers. It could also help spotting scientific misconduct from, e.g., anomalous citation cascades. Dynamic graph clustering is still in its infancy, but it is likely to lead to important technical advances in the next years. A promising approach consists in using the full time-stamped information of the system, instead of focusing on individual subsets corresponding to specific time windows, as it has been done in most of the existing literature. 

The clusterings obtained from the different systems could be integrated in dedicated platforms, in order to limit the biases coming from the individual systems (citation and collaboration habits, for instance) and have a more comprehensive picture.

Which properties should such platforms possess? What consequences would they have for current practices regarding ownership and copyright? How susceptible are these systems to hyping or herding? How to guarantee the longevity of the scientific discourse that is increasingly taking place on these platforms (instead of in the traditional conferences and their proceedings)?

\subsubsection{Structuring of Publications and Citation Objects}
Many of the innovations sketched above are seriously hampered by the current form and format of the standard academic publication: the PDF file. There is a good reason why one of the trend-setting workshops in early 2011 on innovating science was called "Beyond the PDF". Obviously, scientific papers are rich in internal structure: motivations, related work, statement of hypothesis, description of method, interpretation of results, and many other parts are typically found in scientific papers. Furthermore, these parts are linked through a rich rhetorical structure: the motivation justifies the choice of hypothesis, result-interpretation falsifies the hypothesis, the future work builds on comparing the original motivation to the actual results obtained, etc. The only justification for publishing this rich structure as a single monolithic block is historic limitations of printing and book-binding technologies, along with limitations of a low frequency postal system. But now that information technology has freed us from these limitations, can we instead publish our scientific results in a richer format? Can we break up the PDF into its constituent parts, can we make explicit (and typed) links between these parts, can we cite such separate parts, Can we construct (either manually or automatically) rhetorical structure graphs which allows us visualize and analyze the debate that goes on in a scientific community?

\section{Expected Impacts}
\label{sec:imp}
The implementation and wide adoption of the proposed Innovation Accelerator will have a major impact not only on how science is conducted but also how it is used and understood.

\subsection{Impact on Science}
Science and technology today are global and need to be studied, understood, managed, and accelerated globally, i.e., internationally and interdisciplinary. The proposed science accelerator has the potential, we believe, to advance our understanding of the structure and dynamics of science and technology. It is expected to be unique in data coverage and quality as well as the technical sophistication of analysis and visualization components and workflows applied. Massive data streams of papers, patents, funding, news, job advertisements, and relevant social media will be harvested. Algorithms developed in economics, social science, information science, physics, etc. will be applied to mine this data. Graphic design and information visualizations will be employed to communicate the structure and dynamics of science to a diverse audience. While prior work has addressed diverse parts of the S\&T enterprise, the science accelerator will exploit synergies among many parts whose integration is more than their sum.

Based on a systemic understanding of S\&T, we will implement technology that speeds up and improves the conduct of science, the development and refinement of technology, and revenue generation across all S\&T sectors. Among others, the tools and services will make it possible to identify perverse incentives that encourage conservatism, herding and hyping \cite{borgman2012conundrum}; to determine relevant expertise for hire, reviews, or editorial work; to dynamically form (interdisciplinary) productive teams; to effectively ÒmarketÓ research results to gain a higher ÒincomeÓ in terms of download counts and citations; to identify emerging areas of research; and to decrease product development and commercialization. 

The science accelerator will break up the monolithic paper unit, e.g., into nano-publications and interlink them with data, tools, experiment workflows, resources in support of novel search, e.g., retrieve all experts or papers that used a specific dataset or algorithm. It will apply semantic web technology to semantically encode S\&T data in machine-readable format in support of advanced mining and reasoning. 

In a not so far future, the science accelerator will empower many to conduct high quality S\&T with full access to existing knowledge, data and tools and the power to diffuse results effectively to other scientists, educators, funding agencies, policy makers, students or industrial innovators.

\subsection{Impact on Competitiveness}
TodayÕs Òknowledge societyÓ and tomorrowÕs ÒInnovation economyÓ increasingly rely on efficient access to global knowledge, technology, and expertise. Today, Web2.0-style tools help localize and evaluate the key persons, ideas, publications, patents, as well as tangible and intangible resources that are critical for RTD (research and technological development) and other forms of innovation. Many of these services are in the hands of individual providers, and while there is a potential to leverage such information to boost individual, institutional or even country-level innovation, there are strong limits to this process as well. Limitations may in part be caused by the proprietary nature of information, but even more importantly because of the lack of a coherent support system specifically targeted to innovation-seekers. IA as described in this paper envisages putting similar information services into the public domain and focusing these on the very innovation process itself. 

A European vision (amplified in the currently approved ÒHorizon 2020Ó programme) is to turn Europe into an Innovation Union and we strongly believe that the IA initiative can be a cornerstone of this strategy. The innovative methods and tools to be developed and deployed here have the potential to re-invent, if not revolutionize innovation. Different stakeholders can benefit from this in different ways.

\vskip 10pt
\noindent
\textbf{A Head Start for Industry.} Industry will be able to use the inventive IA tools to find expertise, scientific results, and collaborators more effectively, leading to a speed up in ÒR\&D translationÓ time. This way, saving on time and expenditures, a fast and quick moving innovation interface can be established for European companies. 

\vskip 10pt
\noindent
\textbf{Shortcutting the Innovation Cycle.} Classic innovation is a multi-step process with multiple contexts that are different for research, development, and industry. The IA envisions putting these into a common platform, thereby establishing an interchange via a common context that may help the goals of the new European framework in order to, e.g., to form new private-public partnerships, which can create new services to the benefit of society.

\subsection{Impact on Society }
A key aspect of many of the innovations that have been proposed in this paper is that they aim making the innovation chain more transparent: Linked Open Data, publication of scientific data and workflows, open transparent reviewing procedures and tools for trend detection (to name just a few) are all aimed at increased transparency of the innovation system as a whole and of the science system in particular.
Besides the obvious overall impact on society through increased competitiveness, an innovation accelerator as sketched in this paper would also have other direct impacts on various important parts of society.

\vskip 10pt
\noindent
\textbf{CitizensÕ Access to the Innovation Cycle.} Currently, access to the innovation cycle is limited to academia and its spin-off companies, to government (to a limited extent) and to large and often multi-national corporations.  Access to the innovation cycle (and certainly to the early steps in this cycle) is much harder for parties such as local government, lobbying groups (large ones such as Greenpeace, but also smaller ones), trade unions, NGOs, citizen's organizations such as neighborhood committee's and even individual citizens.  Tools such as Linked Open Data, publication of scientific data and workflows, and tools for trend detection would allow these other parties to gain early access to the innovation cycle, provided that such tools are equipped with appropriate user-interface technology.

\vskip 10pt
\noindent
\textbf{Citizens' Influencing the Innovation Cycle.} Currently, citizens are limited to being the passive receivers of the outcomes of the innovation cycle. They have only very limited and indirect means of influence at their disposal, through consumer behavior or through the political system.  Early access to the innovation cycle would also enable these organizations to not only observe but also influence the activities inside the innovation cycle more directly, and ensure that the innovations are to their benefit and that they can influence the trade-offs typically involved in new technologies.

\vskip 10pt
\noindent
\textbf{Education's Access to the Innovation Cycle.} Finally, and perhaps most obviously, the education system would stand to benefit from more transparency in the innovation cycle.  Certainly university education, but also post-academic "life-long learning" and even high-school education would be able to use realistic examples from the participants in the innovation cycle, and would even be in a position to contribute. The broad and increasingly popular idea of "citizen science" is crucially dependent on transparency in the innovation cycle.

\subsection{Synergies with Other Exploratories}
The Innovation Accelerator cannot exist in isolation, but will be embedded into the fabric of science and technology in Europe. Strong synergies with other parts of the overall FuturICT effort (\url{http://www.futurict.ethz.ch}) are expected. Specifically, the Innovation Accelerator is envisioned to interconnect the Exloratories of Economy \cite{cincotti2012}, Technology \cite{batty2012,majsan2012}, Society \cite{alex2012}, and Environment \cite{deffuant2012}.

\begin{figure}
	\centering
	 \includegraphics[width=0.95\linewidth]{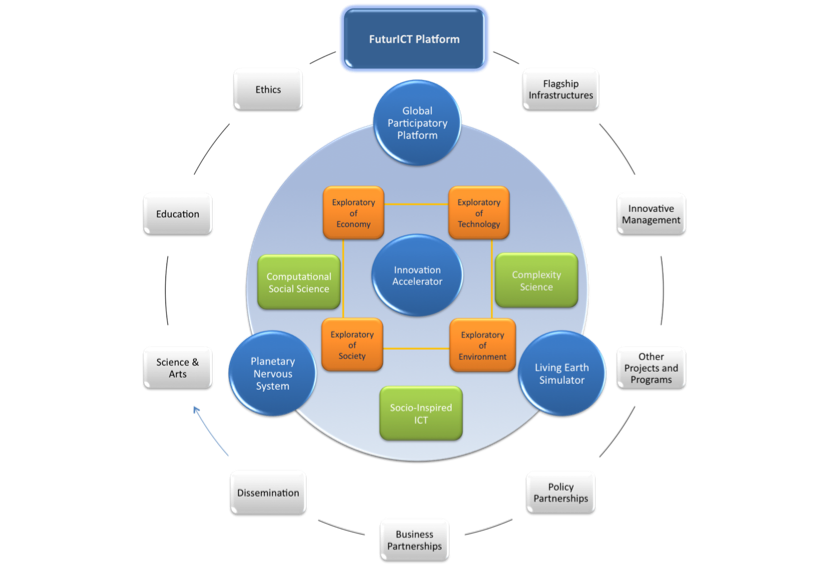}
	 \caption{Main components and activity areas of the FuturICT flagship project.}
	\label{fig:iafict}
\end{figure}

In the programmatic document "FuturICT -- New Science and Technology to Manage Our Complex, Strongly Connected World" by Dirk Helbing \cite{helbing2011fict} (see also \cite{helbing2012} in this volume), a key figure depicts the Innovation Accelerator as occupying a central position in the plan, serving and supporting the various FuturICT components (Figure~\ref{fig:iafict}). Further, FuturICT endeavours to concentrate on "Grand Scientific Challenges" \cite{helbing2012a}, where access to quality information, the organization of research partners, and a rapid, timely access to scientific data and metadata will be critical - in other words, all areas where IA promises to offer a breakthrough.

\section{Discussion}
The present publication is an "ideas paper", one that endeavors to map and explore a territory; as such, it occupies a ãmiddle groundÓ between the original White Paper on the Innovation Accelerator \cite{helbing2011create}, and the more technical, development-oriented papers yet to come, that should elaborate and at the same time narrow down the many topics highlighted here. This is expected to happen in the FuturICT flagship project, in a strong international and inderdisciplinary collaboration, as a high gain -- high visibility enterprise. The goal of the present paper is thus twofold: to persuade the reader of the desirability and feasibility of an IA and to pave the road to further developments.

It is important to note that we do not suggest the immediate and universal revision of the existing science system: before our suggestions (and indeed any similar suggestions) would be upscaled, they have to be thoroughly tested, and this is indeed the meaning of our proposal -- by pursuing the IA idea, to test the feasibility of new reputation, incentive, review and publishing systems and to mature the idea thereby. It is sometimes not easy to improve on the science system, but it is certainly possible. It is, we maintain, worthwhile to extend the existing methods and practices by introducing new varieties. It is clear that the organization of quality control remains central to any reform and several careful steps will be necessitated before a general change is anticipated. This is, in fact, quite typical in science (Thomas S. Kuhn called it the "essential tension" between value preserving and progress, or tradition and innovation \cite{kuhn1979essential}). Instead of a unitary, encompassing framework, we thus expect the emergence of useful experimentation that we would like to endorse. The context of the envisioned FuturICT flagship assures that our vision may be more than just another utopia: in that framework, a coordinated effort will be devoted to the design, implementation and (publisher- as well as project-level) exploitation of the IA.

There are many further issues to be elucidated, for example the new publication items suggested (nanopublications, workflows, data sections of regular papers) will eventually overflow the system with information, so it is crucial that all actors have a way of getting directions in this ocean of scientific bits. We mentioned that such recommendation systems are going to be a relevant component of the IA. This as well as the many other open facets of the IA vision underline the importance of experimentation -- setting the IA ideas to work will produce a better understanding and will help identifying (and solving) new problems. The next step is thus expected to be a \emph{ÒDesign for an Innovation AcceleratorÓ}.

\subsection*{Acknowledgments}
\label{ack}
{\small 
The publication of this work was partially supported by the European Union's Seventh Framework Programme (FP7/2007-2013) under grant agreement no.284709, a Coordination and Support Action in the Information and Communication Technologies activity area (`FuturICT' FET Flagship Pilot Project).}

\bibliographystyle{unsrt}
\bibliography{IA_paper}

\end{document}